\documentclass[12pt,preprint]{aastex}

\newcommand{\sgras}{Sgr~A$^{\ast}$}
\newcommand{\dsos}{$\Delta S/S$}
\newcommand{\sm}{\hbox{$M_{\odot}$}}

\slugcomment{Accepted for publication in ApJL}

\shorttitle{Intra-day Variation of \sgras\ at mm-Wavelengths}
\shortauthors{Miyazaki et al.}

\begin{document}

\title{Intra-day Variation of Sagittarius~A$^{\ast}$ 
at Short Millimeter Wavelengths}

\author{Atsushi Miyazaki}
\affil{Nobeyama Radio Observatory
\footnote{Nobeyama Radio Observatory (NRO) is a branch of 
the National Astronomical Observatory, 
National Institutes of Natural Sciences, JAPAN. }, 
National Astronomical Observatory of Japan, \\
Minamimaki, Minamisaku, Nagano 384-1305, Japan}
\email{amiya@nro.nao.ac.jp}
\author{Takahiro Tsutsumi}
\affil{National Astronomical Observatory of Japan, 
2-21-1 Osawa, Mitaka, Tokyo 181-8588, Japan}
\email{tsutsumi@alma.mtk.nao.ac.jp}
\and
\author{Masato Tsuboi}
\affil{Nobeyama Radio Observatory, 
National Astronomical Observatory of Japan, \\
Minamimaki, Minamisaku, Nagano 384-1305, Japan}
\email{tsuboi@nro.nao.ac.jp}

\begin{abstract}

We have performed the monitoring observations of flux density 
of Sagittarius~A$^{\ast}$ 
at short millimeter wavelengths (100 and 140~GHz bands) 
on seven years in the period from 1996 to 2003 
using the Nobeyama Millimeter Array (NMA). 
We found intra-day variation of \sgras\ in March 2000 flare.  
The flux density at the peak of the flares increases 100--200\% 
at 100~GHz and 200--400\% at 140~GHz (\dsos), respectively.  
The two-fold increase timescale of the flare 
is estimated to be about 1.5 hours at 140~GHz.  
The intra-day variation at mm-wavelengths 
has similar increase timescale as those in the X-ray and infrared flares 
but has smaller amplitude.  
This short timescale variability suggests that 
the physical size of the emitting region 
is smaller than 12~AU ($\approx 150~R_{\rm s}$).  
The decay timescale of the flare was at most 24 hours.  
Such a light curve with rapid increase and slow decay is similar to 
that often observed in outburst phenomena with ejections.  

\end{abstract}


\section{Introduction}

Sagittarius~A$^{\ast}$ (\sgras) is the compact radio source 
which is located at the dynamical center of our galaxy 
and believed to be associated with the supermassive black hole 
of a mass of about $4 \times 10^6~\sm$ 
(e.g., Ghez et al. 2003; Sch\"{o}del et al. 2002, 2003).  
Because this source is embedded in thick thermal material, 
it is practically difficult to observe its fine structures 
by the present VLBI (very long baseline interferometry) 
(Doeleman et al. 2001; however, see also Bower et al. 2004, 
the recent detection of the intrinsic size of \sgras\ using VLBI technique).  
Time variability observation is a powerful and alternate tool 
to probe the structure and the emission mechanism of \sgras.  
If the time variability is intrinsic in the source, 
it should be tightly related to 
the structure and the emission mechanism of the emitting region. 
If the origin is not intrinsic, the observation should at least provide 
the information about thermal materials around \sgras.  

The time variability of \sgras\ at cm-wavelengths 
has been studied in these two decades \citep{bl82,zha01}.  
The variability at short cm-wavelengths seems to have a periodicity 
of about 106 days \citep{zha01}.  
It is, however, an open question what causes the 106-days cycle.  
On the other hand, the variability at millimeter wavelengths 
has not been well observed 
although it believed to be caused by the activity of \sgras\ itself.  
Earlier, \citet{wb93} reported that there were 
significant flux variations of \sgras\ at 86~GHz in several tens of days.  
The extractable information from such observations 
has been limited because of sparse observation intervals.  
Thus we performed the systematic flux monitor of \sgras\ 
at 100 and 140~GHz bands in winter seasons of 1996, 1997, and 1998 
using Nobeyama Millimeter Array (NMA) 
which is a six 10-m dish element interferometer 
at the Nobeyama Radio Observatory \citep{mtt99,tmt99}.  
The flux density of \sgras\ 
was measured every several days for two months.  
In these monitor observations we found a flare of \sgras\ 
and showed that the timescale of the flare 
is at most a several days \citep{mtt99,tmt99}.  
The variability amplitudes at mm-wavelengths 
are larger than those at cm-wavelengths.  
Similar flares of \sgras\ have also been detected 
at 1~mm wavelength \citep{zha03}.  
On the other hand, 
X-ray flares of \sgras\ were recently detected 
by {\it Chandra} \citep{bag01} 
and {\it XMM Newton} \citep{por03,gol03} observations.  
For example, the flare detected by \citet{por03} is rising in a few ks and 
with a peak luminosity $\sim$160 times higher than the quiescent state.  
The short timescale of these flares shows 
that the emission arises from the region quite close to the \sgras.  
Recently, infrared flares were also detected \citep{ghe04,gen03}.  
The flare detected at at H, K$_{\rm S}$, and L$'$-bands 
has a factor of about 2--4 variability 
on timescale of several tens of minute \citep{gen03}.  
The infrared flares are similar to the X-ray flares 
in the duration, rise/decay times and band luminosities \citep{gen03}.  
The apparent correlations 
between radio and X-ray flares \citep{zha04}, 
between infrared and X-ray flares \citep{eck04}, 
and between radio, infrared, and X-ray flares \citep{bag02} 
were also reported.  
The relations among X-ray, infrared, and radio flares 
should be important for probing the emission mechanism of \sgras.

We have conducted intensity monitoring observations 
toward \sgras\ at short mm-wavelengths using  NMA.  
In this paper, we will concentrate on intra-day variation of \sgras.  
The long time variability will be presented in another paper.  
In \S 2, we present the detail of observations and calibrations.  
In \S 3, we present the results of monitoring observations 
and discuss the properties of flares.  
In this paper, we assume that the Galactic center distance 
is 8.0~kpc \citep{rei93}.

\section{Observations and Calibrations}

We have performed intensity monitoring observations toward \sgras\ 
at 100 and 140~GHz band ($\lambda$=3 and 2~mm) 
using the NMA since 1996 (also see Tsuboi et al. 1999; Miyazaki et al. 2003).  
The observations were carried out over winter to spring.  
Each epoch consists of a set of sequent observations of 2--3 days.  
The epochs of the observations in 2000 were separated by about 5 days.  
The observational dates in March 2000 are summarized in Table~\ref{tbl-1}.  
The observing time for each day was 2--3 hr.  
The maximum observable time of \sgras\ with NMA is 4 hr.

\begin{figure}
\epsscale{0.8}
\plotone{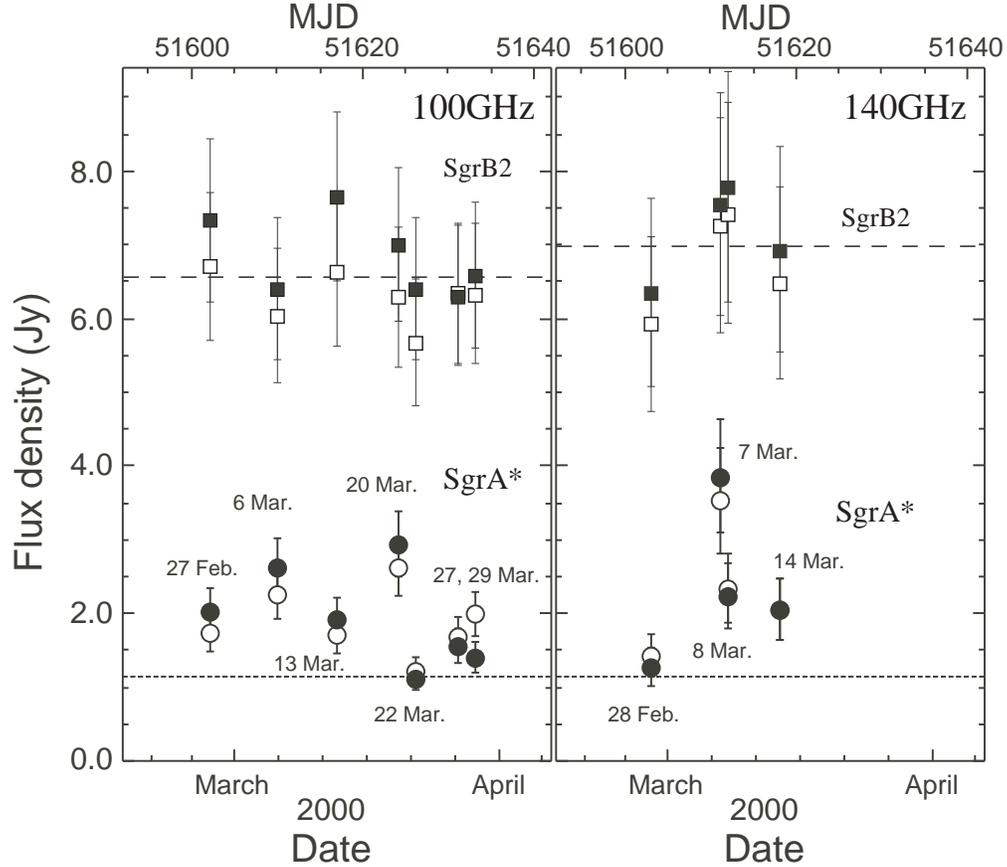}
\caption{The light curve of \sgras\ (circles) 
at 100 (left panel) and 140~GHz (right panel) bands 
during March--April 2000.  
Open and filled circles indicate the observed frequencies, 
90 and 102~GHz for 100~GHz band, 
and 134 and 146~GHz for 140~GHz band, respectively.  
The flux density at 100~GHz was violently changing.  
There is a steep peak at 140~GHz band on 7 March.  
The flux densities of \sgras\ at the peak 
were $3.5\pm0.7$~Jy at 134~GHz 
and $3.9\pm0.8$~Jy at 146~GHz, respectively.  
The horizontal dotted line indicates the mean flux density 
in a quiescent phase.  
Moreover, squares indicate 
the measured peak flux density of Sgr~B2(M) (see \S 2).  
The horizontal dashed line indicates the mean flux density of Sgr~B2(M), 
6.6~Jy at 100~GHz band and 7.0~Jy at 140GHz band, respectively.  
The scatters of the flux densities 
are less than the absolute uncertainties.  
\label{fig:all00}}
\end{figure}

\sgras\ was simultaneously observed in two frequencies, 
either at 90 and 102~GHz or 134 and 146~GHz 
using both lower and upper side bands each 
with a bandwidth of 1~GHz \citep{oku00}.  
We observed NRAO~530 and 1830-210 
every 20 minutes as phase calibrators.  
The flux densities of these calibrators are determined from 
Uranus or Neptune which were used as the primary flux calibrators.  
The absolute uncertainties of the flux scaling, 
which are mainly caused by phase instability 
and signal-to-noise of phase calibrators, 
are about 15\% and 20\% at 100 and 140~GHz bands, respectively.  
The flux densities of NRAO~530 and 1830-210 at March 2000 were 
2.4~Jy and 2.5~Jy for 100~GHz band, and 
1.8~Jy and 1.9~Jy for 140~GHz band, respectively.  
To correct instrumental and atmospheric effects, 
the time-dependent phase and amplitude gains 
were determined from the phase calibrators 
using the standard NMA data reduction package UVPROC2 \citep{tmu97}.  
The elevation dependence of atmosphere absorption 
is calibrated by system temperature.  
To check the validity of flux calibration, Sgr~B2(M) 
[R.A.$=17^{\rm h}44^{\rm m}10\fs4$, 
Decl.$=-28\arcdeg22\arcmin03\arcsec$ 
(B1950.0)] was also included in the same track 
as the synthesis observations of \sgras\ 
and its flux density was determined 
using the same flux scale as in the case of \sgras.  
The flux densities of Sgr~B2(M) is not variable 
because this is a compact \ion{H}{2} region.  
Figure~\ref{fig:all00} shows the peak flux density of Sgr~B2(M) 
in March 2000 (squares).  
The measured mean flux densities (horizontal dashed line) 
are 6.6~Jy at 100~GHz band and 7.0~Jy at 140GHz band, respectively.  
The scatter of the flux densities shows the absolute uncertainties.  

Almost all observations including the detections of flares 
of \sgras\ were performed by the array configuration 
with intermediate baselines of the NMA.  
The projected baselines 
range $\sim$7-55 k$\lambda$ at 100~GHz band 
and $\sim$10-77 k$\lambda$ at 140~GHz band.  
The visibilities with projected baselines over 25 kilo-wavelengths 
($(U^2+V^2)^{1/2} \ge 25 k\lambda$) are used 
in order to suppress the contamination 
from the extended components surrounding \sgras.  
We CLEANed the maps with the restricted visibilities 
using the AIPS package.  
Typical synthesized beam sizes (HPBW) 
were about $3'' \times 6''$ and $2'' \times 4''$ 
at 100 and 140~GHz bands, respectively.  
The observed flux density is reduced by the phase error 
due to atmospheric phase fluctuation of timescales less than 
the scan interval of the phase calibrators.  
From the flux densities of the calibrator measured on the maps, 
the fractions of the decorrelation are 
10--20\% for 100~GHz band and 20--40\% for 140~GHz band.  
In order to correct the decorrelation, 
the observed flux densities of \sgras\ were divided by the correction factors, 
0.8--0.9 for 100~GHz band and 0.6--0.8 for 140~GHz band, respectively.  
We averaged two measured flux densities of \sgras\ 
which were individually calibrated by the two phase calibrators.  
After this calibration process, 
the residual contribution from the extended components 
in the flux measurements is smaller than 
0.2~Jy and 0.1~Jy at 100 and 140~GHz band, respectively.

\section{Results}

A total number of observations of \sgras\ at 100~GHz band 
from 1996 to 2003 is about 60 days.  
The light curve shows that \sgras\ 
has quiescent and active phases \citep{mtt03}.  
Mean flux densities in a quiescent phase 
are $1.1\pm0.2$~Jy and $1.2\pm0.2$~Jy at 90 and 102~GHz, 
respectively.

\begin{figure}
\epsscale{0.7}
\plotone{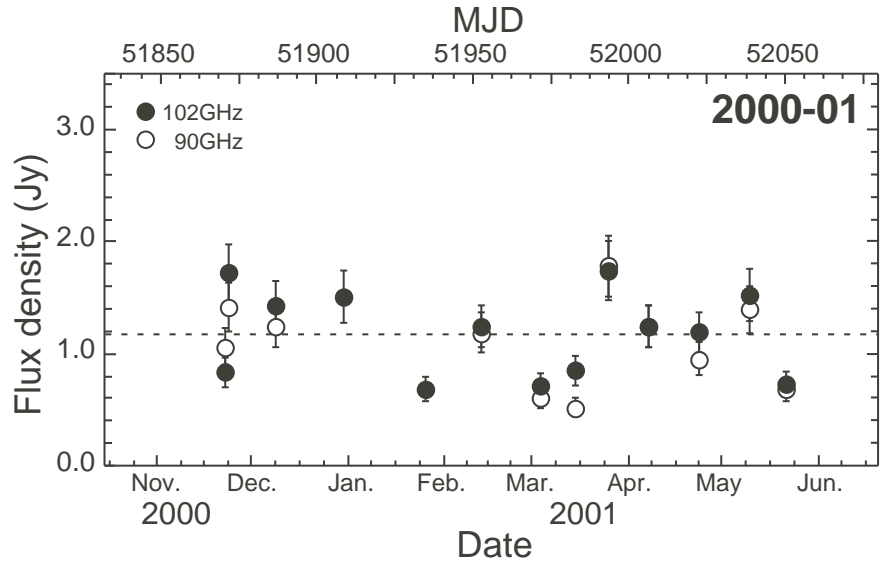}
\caption{The light curve of \sgras\ at 100~GHz band 
during November 2000--May 2001.  
\sgras\ was probably in a quiescent phase.  
The horizontal dashed line indicates the mean flux density 
in a quiescent phase.  
\label{fig:qui}}
\end{figure}

Figure~\ref{fig:all00} shows the light curves of \sgras\ 
at 100 and 140~GHz bands in 2000.  
This is a representative of the active phase.  
Figure~\ref{fig:qui} shows the light curves of \sgras, 
which is probably the quiescent phase, at 100~GHz in 2000--2001.  
The flux densities in the figure were averaged for one observation day.  
There was violent variability in the active phase in 2000.  
Several flares with durations of days to a few weeks were identified.  
A most prominent flare was observed on 7 March 2000 at 140~GHz band.  
The peak flux densities of \sgras\ at 134 and 146~GHz 
were $3.5\pm0.7$~Jy and $3.9\pm0.8$~Jy, respectively.  
The flux density then decreased to $2.2\pm0.4$~Jy at 146~GHz 
on the subsequent day, 8 March 2000.  
Weather condition on 7 and 8 March was fine.  
The water vapor pressure was less than 2~hPa, 
which is translated to opacity $< 0.1$ at zenith, 
during the observations on both 7 and 8 March.  

The half decay timescale of the flare at 146~GHz, was at most 24 hours.  
The averaged quiescent flux was about 1~Jy at 140~GHz band.  
The flare amplitude was about 300\% (\dsos) 
of the mean flux density level at 146~GHz, 
which is larger than that at 100~GHz band (200\%).  
This probably indicates that the variability increases with frequency.

\begin{figure}
\epsscale{0.6}
\plotone{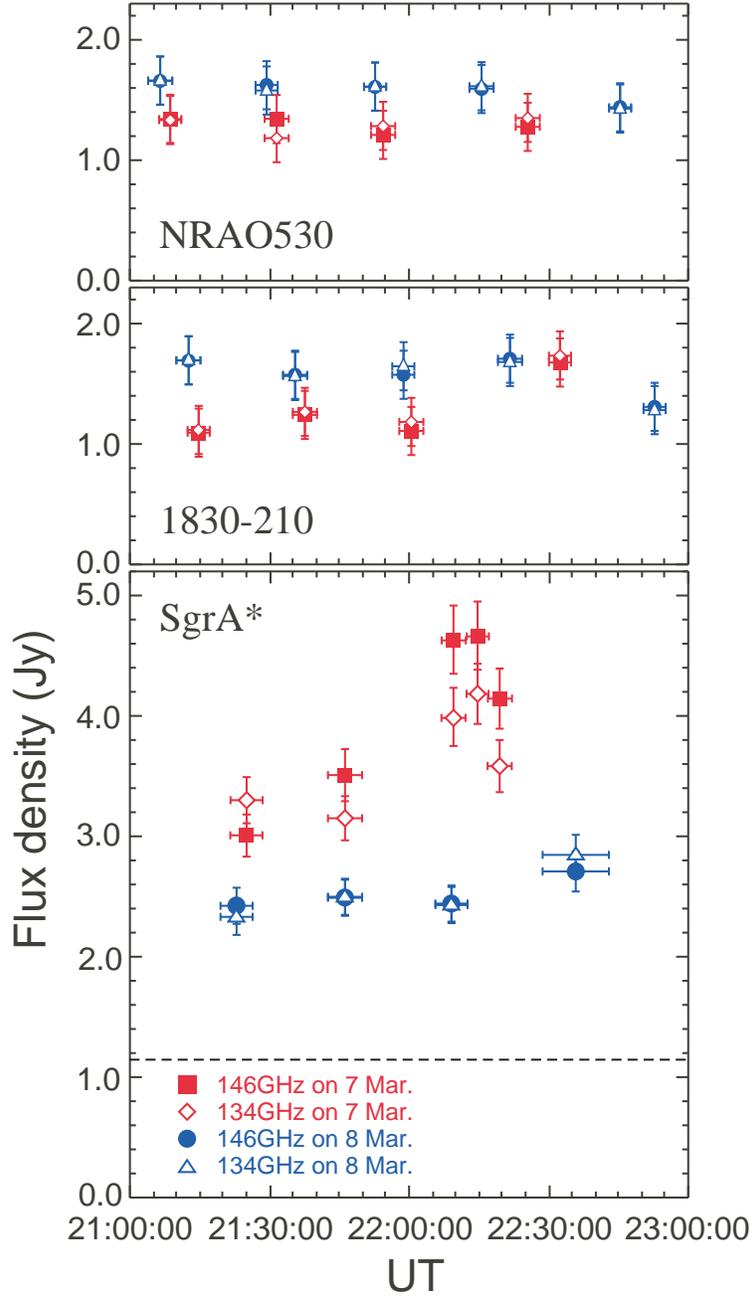}
\caption{The light curve of the \sgras\ flare (bottom panel) 
and the calibrators (upper panels), NRAO530 and 1830-210, 
at 140~GHz band on 7 (in red) and 8 (in blue) March 2000.  
Filled squares, open diamonds, filled circles, and open triangles 
indicate the observed frequencies and dates, 
146 and 134~GHz on 7 March, and 146 and 134~GHz on 8 March, respectively.  
The flux density of \sgras\ at 146~GHz (red filled squares in bottom panel) 
increased from 3.5 to 4.7~Jy from 21:45 to 22:15 UT on 7 March.  
The horizontal dashed line indicates the mean flux density at 140~GHz band.  
\label{fig:idvm00}}
\end{figure}

We divided the data set at 140~GHz band observed on 7 and 8 March 2000 
into about a 5 minute bin around the peak and 7--14 minute bins for others 
and measured the flux density of \sgras\ at each bin 
in order to search for shorter timescale variability.  
Figure~\ref{fig:idvm00} shows the light curve of \sgras\ 
at 140~GHz in the two days.  
On 7 March, the flux density around the peaks changed rapidly.  
These are summarized in Table~\ref{tbl-2}.  
The peaks of the flares at 134 and 146~GHz were occurred at 22:14 UT.  
The flux density of \sgras\ at 146~GHz 
increased from 3.5 to 4.7~Jy between 21:45 to 22:15 UT on 7 March.  
The peak flux densities were 
$4.2\pm0.8$ Jy at 134~GHz and $4.7\pm0.9$ Jy at 146~GHz.  
We used the same flux scales, which was determined 
from the flux calibration in March 2000, 
for both data on 7 and 8 March.  
The relative uncertainty in the two days depends only on 
the accuracy of the gain calibration and the estimation of decorrelation.  
Then the relative uncertainty is smaller than the absolute one.  
The typical relative uncertainties 
of 100 and 140~GHz bands 
were estimated to be a few \% and 6\%, respectively.  
The upper panel in Figure~\ref{fig:idvm00} shows the light curves 
of the calibrators (NRAO530, 1830-210).  
The scatters of the measured flux density are much smaller 
than the variation in the flare of \sgras.  
The projected baselines change with earth rotation.  
This change in sampling of the source structure 
might cause artificial variation in flux density.  
However, such effect is not significant 
because no variation in flux density 
was observed within the relative uncertainty on 8 March.  
Thus the observed 30\% increase in 30~minutes must be real.  
The timescale that the flux density increased by 100\% 
(two-fold increase timescale) 
is estimated to be about 1.5 hours 
assuming that the increase has a constant gradient.  
On the other hand, the intra-day variation 
was not found in the 100~GHz band data taken 
during the active phase in March 2000.  
The intra-day variation of \sgras\ at cm-wavelengths 
reported by \citet{bow02} indicates that 
the 15~GHz flux density increased by about 10\% in 2 hours.  
The amplitude of the variation in our mm-wavelength observations 
is much larger than the value of the cm-wavelength observations.  
The intra-day variation at mm-wavelength has been also 
reported for the galactic nucleus of M81 \citep{sak01}.

\section{Discussion}

The peak flux at 146~GHz corresponds to a factor of 4.5 
increases from the mean value in the quiescent phase.  
The radio luminosity of the flare at the peak 
is $L_{\rm R} \approx 3 \times 10^{34}$~erg$\:$s$^{-1}$ 
assuming that the frequency width of the flare is 150~GHz.  
The X-ray flare is rising at about 1 hour 
and with the peak luminosity observed by {\it XMM-Newton} of 
$3.6^{+0.3}_{-0.4} \times 10^{35}$~erg$\:$s$^{-1}$ \citep{por03}.  
On the other hand, 
the peak luminosity of infrared flares 
observed by the Very Large Telescope at {\it H}-band is about 
$9 \times 10^{35}$~erg$\:$s$^{-1}$ \citep{gen03}.  
The observed infrared flares are similar to the X-ray flares 
in the duration, rise/decay times and luminosities \citep{gen03}.  
The flare at mm wavelengths has similar increasing timescale 
as the X-ray and infrared flares though has smaller amplitude.  

During the flare peak, flux densities at 146~GHz 
became larger than those at 134~GHz.  
The last column in Table~\ref{tbl-2} shows 
spectral indices, $\alpha$, estimated between 134 and 146~GHz 
($S_\nu \propto \nu^\alpha$).  
The positive spectral index during the flare peak is clear 
though an uncertainty in the index is large due to a small frequency span.  
As shown in Figure~\ref{fig:all00}, the peak flux density 
at 140~GHz band are also larger than that at 100~GHz band.  
This demonstrates that the spectrum was steep during the flare phase.  
The steep spectrum was also observed in the flare 
observed in 1998 \citep{tmt99}.  
The spectral variation suggests that the energy injection to photons 
was occurred in the higher frequency regime and 
the emitting frequency was shifted to millimeter wave regime with time.

The increasing timescale of the mm flare, 1.5 hours, 
provides the physical size (light crossing size) of the flare region 
in the accretion disk that is compact 
at or below $\sim 12$~AU ($\approx 150~R_{\rm s}$; 
Schwarzchild radius, $R_{\rm s} \equiv 2GM/c^2$, 
assuming a black hole mass of $4 \times 10^6~\sm$).  
This upper limit is larger than the physical size 
of \sgras\ itself ($\sim 1.4$~AU) observed by VLBI \citep{doe01,kri98}.  
From the upper limit, the brightness temperature at the flare 
is estimated to be higher than $8 \times 10^9$~K.  
There is asymmetry in the light curve as the half decay timescale 
is much longer than the flare increase timescale.  
Such light curve with rapid increase and slow decay 
is often observed in outburst phenomena with ejections, 
for example, flares on the Sun, 
the 1972 outburst of Cyg~X-3 \citep{mar92}, etc.  
The total energy of the flare is in the order of $10^{39}$~erg 
if the duration of the flare is assumed to be a few days.  
The decay rate decreased significantly 
between 8 March and 14 March 
as the flux density on 14 March was 2~Jy, 
which is still twice as large as the mean value in a quiescent phase.  
Then, the total energy of the flare is as large as $10^{40}$~erg.  
However, apparent flattening of the decay rate 
may be due to another unseen flare.

\section{Summary}

We have performed the monitoring observations of the flux density 
of \sgras\ at 3~mm (100~GHz) and 2~mm (140~GHz) bands 
using the Nobeyama Millimeter Array (NMA) 
from 1996 to 2003.  
We detected several active phases.  
In the March 2000 flare, 
the flux densities of \sgras\ at 140~GHz band 
had reached a peak, $\sim$4.5~Jy, on 7 March 
and increased \dsos$\gtrsim 400$\%.  
Then the flux was decreased to a half in a day.  
Moreover, We detected the 30\% flux increase in 30 minutes 
on 7 March 2000.  
The timescale that the flux density increased by 100\% 
is estimated to be about 1.5 hours 
assuming that the increase has a constant gradient.  
The upper limit for a size of the variable component 
estimated from the timescale of this intra-day variability 
is a few tens of AU.  
Such a light curve with rapid increase and slow decay 
is similar to that often observed in outburst phenomena with ejections.

\acknowledgments

We thank the staff of NMA group of the Nobeyama Radio Observatory 
for support in the observation.



\clearpage
\begin{deluxetable}{rcc}
\tablecaption{Observational dates of \sgras\ in 2000 \label{tbl-1}}
\tablewidth{0pt}
\tablehead{
\colhead{Date} & \colhead{Time (UT)} & \colhead{Frequency (GHz)}}
\startdata
February~27 & 21:10-23:00 & 90 \& 102 \\ 
February~28 & 20:50-22:50 & 134 \& 146 \\
March~~6 & 20:50-23:00 & 90 \& 102 \\
March~~7 & 21:10-22:40 & 134 \& 146 \\
March~~8 & 21:00-23:00 & 134 \& 146 \\
March~13 & 20:10-23:00 & 90 \& 102 \\
March~14 & 20:20-23:00 & 134 \& 146 \\
March~20 & 19:40-21:30 & 90 \& 102 \\
March~22 & 19:40-21:30 & 90 \& 102 \\
March~27 & 19:10-21:00 & 90 \& 102 \\
March~29 & 19:10-21:00 & 90 \& 102 \\
\enddata
\end{deluxetable}

\clearpage
\begin{deluxetable}{rccr}
\tablecaption{Flux Densities for the Flare of \sgras\ 
at 140~GHz band in 2000 \label{tbl-2}}
\tablewidth{0pt}
\tablehead{
\colhead{Time (UT)} & \multicolumn{2}{c}{Flux density (Jy)\tablenotemark{a}} 
& \colhead{index\tablenotemark{b}}\\
\cline{2-3} 
\colhead{} & \colhead{134~GHz} & \colhead{146~GHz} & \colhead{}}
\startdata
2000~March~7~~21:21-21:28 & $3.29 \pm0.2$ & $3.01 \pm0.2$ & $-1.1$\\ 
21:44-21:51 & $3.15 \pm0.2$ & $3.52 \pm0.2$ & $1.3$\\
22:07-22:12 & $3.99 \pm0.2$ & $4.63 \pm0.3$ & $1.7$\\
22:12-22:17 & $4.18 \pm0.3$ & $4.67 \pm0.3$ & $1.3$\\
22:17-22:22 & $3.59 \pm0.2$ & $4.15 \pm0.2$ & $1.7$\\
mean\tablenotemark{c} & $3.53 \pm0.2$ & $3.86 \pm0.2$ & $1.0$\\
\cline{1-4} 
2000~March~8~~21:19-21:26 & $2.33 \pm0.1$ & $2.43 \pm0.1$ & $0.5$\\ 
21:42-21:49 & $2.51 \pm0.2$ & $2.50 \pm0.1$ & $-0.1$\\
22:05-22:12 & $2.43 \pm0.1$ & $2.45 \pm0.1$ & $0.1$\\
22:28-22:42 & $2.85 \pm0.2$ & $2.71 \pm0.2$ & $-0.6$\\
mean\tablenotemark{c} & $2.33 \pm0.1$ & $2.23 \pm0.1$ & $-0.5$\\
\enddata
\tablenotetext{a}{The errors indicate the relative errors 
of the flux densities (see \S 3).  } 
\tablenotetext{b}{The spectral indices, 
$\alpha$ ($S_\nu \propto \nu^\alpha$), 
estimated between 134 and 146~GHz.  
The typical errors of indices are about $\pm1.4$. } 
\tablenotetext{c}{The derived value using all data in a day. } 
\end{deluxetable}


\clearpage

\clearpage

\clearpage

\end{document}